# Context-Aware Emotionally Adaptive Voice Assistants: A Multimodal Framework for Empathetic Human-Agent Interaction

Mr. Tapon Kumer Ray
Vellore Institute of Technology, Amaravati, Andhra Pradesh, India.

Dr. Rajkumar Yesuraj
Vellore Institute of Technology, Amaravati, Andhra Pradesh, India.

Corresponding Author: Tapon Kumer Ray, Email: tapon.22bce20245@vitapstudent.ac.in

*Abstract*— Voice-assistant interruptions tend to be intrusive because existing systems fail to consider the affective state, cognitive load and situational context of the user when deciding when and how to interrupt.Voice-assistant interruptions tend to be intrusive, since existing systems do not consider the affective state, cognitive load or situational context of the user when determining when and how to interrupt. In this paper, EmpathicVA, a closed-loop framework integrating physiological sensing, vocal-affect analysis, contextual modeling and reinforcementlearning interruption policy, is introduced. A hierarchical fusion model involves integrating HRA, EDA, respiration, acousticprosodic features, linguistic embeddings, and contextual cues and computing the probabilities of five affective states. A Double Deep Q-Network selects immediate response, brief or extended delay, empathetic response, or silent mode based on these probabilities, context and interaction history. The multimodal model obtained an accuracy of 92.3% and an F1-score of 0.922 at the macro level on a held-out test set, outperforming the highest accuracy unimodal model by 6.0 percentage points. Comparing the six-week within-subject field study with 48 participants with a baseline and context-only assistants, there was a corresponding increase in satisfaction, trust, and appropriateness of timing, as well as a large reduction in interruption-related stress episodes. The results suggest that affect-aware timing and restraint are both important in voice interaction in addition to the response wording.

## I. Introduction

Voice assistants (VAs) have come a long way in their ability to recognize speech and understand language, but proactive interaction is a challenging HCI problem that remains challenging. Field studies and recent reviews demonstrate a strong dependence of the suitability of proactive behavior on user activity, state, setting, and agency, with mixed results outside safety-critical contexts [1]–[4]. Inappropriate, or even premature, interventions can therefore contribute to the sense of intrusion and loss of trust.

Proactive voice-assistant timing has been explored in recent research [1, 3, 4]; multimodal affect recognition has been explored in recent research [7, 9]; and empathic conversational-agent design has been explored in recent research [5, 6]; and RL-based dialogue policy learning has been explored in recent research [13]. However, these lines of research tend to be kept apart. Context-aware systems typically do not require an affect estimate from the system to be calibrated, but rather sense observed activity; Emotionrecognition systems often optimize the classification without any linkage to the interaction policy; Adaptive dialogue systems seldom incorporate physiological costs for intrusive timing.

EmpathicVA makes a contribution to this integration gap in four ways. The first is the introduction of a real-time perception–decision– action architecture that incorporates affect, context and interaction history. Second, it builds a hierarchical multimodal affect model which combines situational, linguistic, vocal, and physiological information. Third, it represents the policy of interruption management as a Double DQN policy in which the action consists of a combination of timing and emotional register control. Fourth, it assesses the framework on held-out multimodal data and a 48-user, 6-week field deployment of VA users. The main research question is whether affect estimation can be reliably transferred to interaction policies that can measurably decrease intrusiveness while maintaining usefulness.

## II. Related Work

### A. Context, Interruptibility, and Voice Interaction

Recent studies on models of proactive VAs have adopted the activity, location, task, social setting and user state as the model for context [1] [3]. The field evidence for voice-based notifications also suggests that good delivery times depend on the activity or context of the device [4]. But, the same observable context can be related to different affective states: the same person (at a desk) can be relaxed, frustrated, or in an acute stress state. Therefore, proactive communication cannot be designed as if it were assumed that users consent or intend something [2].

### B. Multimodal Affect Recognition

Speech is rich in prosodic, spectral and linguistic cues, while physiological signals are evidence of autonomic arousal. Recent surveys and experiments indicate that the fusion of heterogeneous channels can be advantageous for robustness but vulnerable to dataset bias, sensor noise, lack of modalities, and cross-subject variations [7]–[9]. The most widely used features in contemporary speechemotion systems are the MFCCs, spectrograms, convolutional/recurrent models, augmentation and the RAVDESS-based evaluation [10, 11]. However, a correctly identified classification does not indicate if the assistant should respond immediately, delay, acknowledge emotion or remain silent.

### C. Reinforcement Learning for Adaptive Interaction

RL provides a framework for optimizing sequential interaction under delayed feedback. A recent survey of taskoriented dialogue policy learning identifies state representation, action design, reward modeling, exploration, and evaluation as central challenges for RL-based conversational systems [13]. EmpathicVA applies a Double DQN implementation to this setting, including affect probabilities in the state and physiological stress changes in the reward.

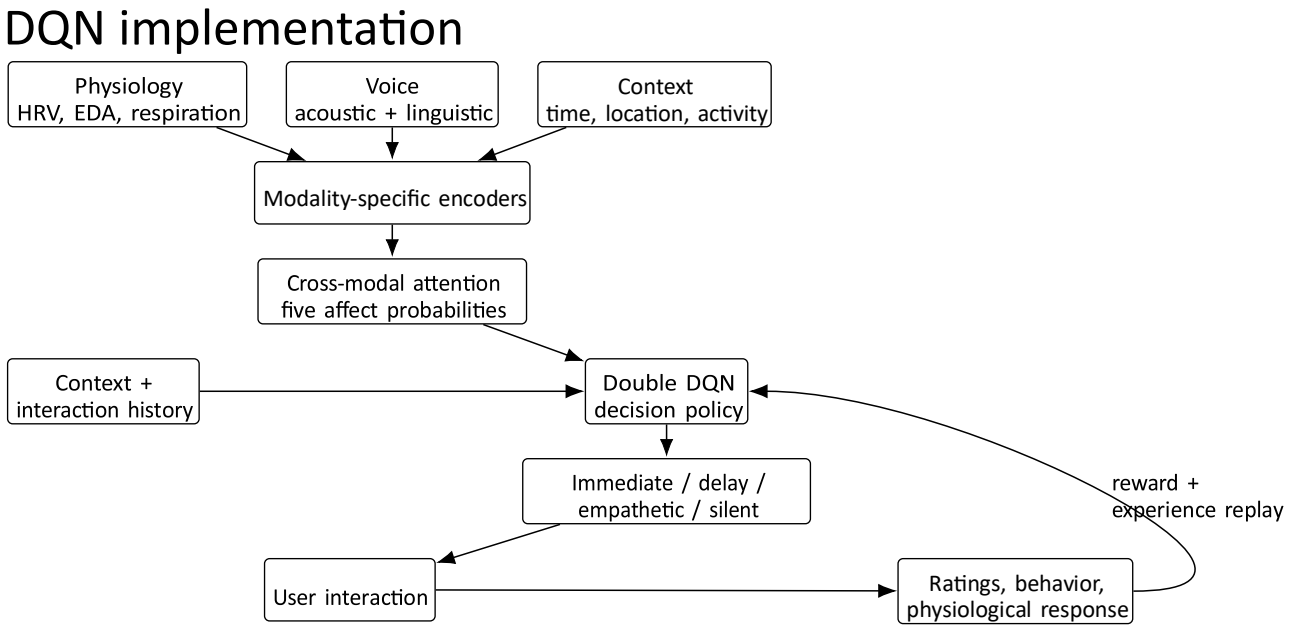

Fig. 1. EmpathicVA perception–decision–action loop.

## III. EmpathicVA Framework

### A. Closed-Loop Architecture

The system is summarized in figure 1. The Multimodal Affect Recognition Module (MARM) is the module that consumes all the physiological, vocal and contextual streams. Each stream is encoded by a modality-specific encoder and the attention across modalities results in a distribution of affects over the 5 classes. The RL agent merges this distribution with the context of situation and past interaction results. Selected action is carried out by the response layer, and explicit ratings, acceptance or dismissal behavior, response latency and post-interruption physiology are returned as feedback.

### B. Multimodal Affect Recognition

HRV, EDA and respiration are processed in 30-s intervals, each overlapping by 15-s, in the physiological pipeline. This fourth order Butterworth filter reduces noise from the sensors, wavelet screening filters out the motion-corrupted part of the data and the subject-specific z-scoring normalisation decreases baseline variation. The HRV features include RMSSD, pNN50, and LF/HF ratio, and the EDA features include tonic skin-conductance level and phasic responses; the respiration features include rate and regularity.

Acoustic and Linguistic evidence are merged in the vocal pipeline.

Acoustic features are comprised of 13 Mel-frequency cepstral coefficients, fundamental-frequency statistics and contours, spectral centroid, bandwidth, roll-off, jitter, shimmer, and harmonicto noise ratio. Language characteristics comprise sentiment polarity, lexical and syntactic features, and BERT-base embeddings. Time and Day are encoded by contextual features, as well as semantic location, active application behavior, notification load, meeting density and estimated free intervals.

Physiological sequences are encoded by a 1-D convolutional layer followed by a 64-unit LSTM. Spectrograms are processed by a 2-D convolutional layer and a 128-unit bidirectional LSTM. A dense context branch encodes the 12 contextual features. Eight-head cross-modal attention dynamically weights the branches; a 128-unit dense layer and five-way softmax output the probabilities of neutral, positive, negative, stressed, and excited states. The model is trained with Adam (learning rate $10^{-3}$, $\beta_1 = 0.9$, $\beta_2 = 0.999$), early stopping with patience 20, stratified five-fold validation, and physiological augmentation using time warping and realistic Gaussian noise. Table I lists the principal model dimensions. The design intentionally separates within-modality temporal modeling from cross-modal fusion: the recurrent layers first learn modality-specific dynamics, while attention is reserved for deciding which modalities are reliable for the current sample. This distinction is important in natural settings because microphone quality, sensor contact, and context availability vary over time.

TABLE I Core Model and Policy Configuration

| Subsystem | Component | Configuration |
|---|---|---|
| Physiology | Conv1D + LSTM | 64 filters; 64 units |
| Voice | Conv2D + BiLSTM | 128 filters; 128 units |
| Context | Dense branch | 12 inputs; 32 units |
| Fusion | Multi-head attention | 8 heads; key dim. 256 |
| Classifier | Dense + softmax | 128 → 5 |
| Policy | Double DQN | 256–128–64–5 |
| Replay | Prioritized buffer | 50,000 transitions |
| Exploration | $\epsilon$-greedy | 1.0 → 0.01 |

### C. RL State, Actions, and Reward

The state is

$$s_t = [p^{\text{affect}}_t, c_t, h_t] \in \mathbb{R}^{37}, \tag{1}$$

where $p^{\text{affect}}_t \in \mathbb{R}^5$ is the affect probability vector, $c_t \in \mathbb{R}^{12}$ is context, and $h_t \in \mathbb{R}^{20}$ summarizes the last ten action–feedback pairs. Retaining probabilities rather than hard labels allows the policy to account for uncertainty.

The action set contains five behaviors: immediate response (within 2 s), brief delay (10–30 s), extended delay (1–5 min), empathetic response, and silent suppression of nonurgent notifications. The reward combines explicit satisfaction, inferred timing quality, affective congruence, and physiological intrusiveness:

$$R_t = .4R_{\text{sat}} + .3R_{\text{time}} + .2R_{\text{emp}} - .1R_{\text{intr.}} \quad (2)$$

Timing quality uses response latency and acceptance or dismissal; intrusiveness uses changes in HRV and EDA after an interruption. The Q-network has 37 inputs, hidden layers of 256, 128, and 64 ReLU units, and five linear outputs. Training uses Double DQN, a 50,000-transition prioritized replay buffer, batch size 64, a target-network update every 1,000 gradient steps, and $\epsilon$-greedy exploration decaying from 1.0 to 0.01.

*D. Execution of Response and Personalization*

The action layer is the response layer, which translates selected actions into observable assistant behaviour. Immediate, brief-delay, and extendeddelay actions are actions that do not change the underlying task response, but do change the delivery time. Empathetic response: Brief emotional acknowledgement prior to content that is relevant to the task; no therapeutic statements. Silent mode will mute only the non-urgent proactive notifications, and will not eliminate any user-initiated commands or safety-critical notifications. This separation enables EmpathicVA to be integrated with an existing VA backend without the need for the language model to be retrained.

Indirect representations of personalization are interaction-history state and repeated observation of rewarding. If the interaction is accepted, for example, it will strengthen the value of immediate response for similar states of the world in the future, but if it is rejected it will weaken it. The ten events are deliberately shorter in duration, but have enough time to encompass recent trends in preference as well as stress buildup. A deployed version would need to be periodically recalibrated due to the fact that affective baselines and schedules change. Conservative fallback behaviour – if there is poor quality of sensor, and/or high uncertainty in affect distribution, the system defaults to context and is more inclined to delay than get emotionally involved in the intervention:

The entire loop is conceived for quasi real-time operation. The physiological windows are updated every 15 s, and the vocal and context features are updated at every utterance and/or relevant device event. The affect model then provides a probability vector to the policy. Since the decisions are to be made during interruption, the small DQN does not dominate the inference latency, which is mostly determined by feature extraction. Local processing also decreases reliance on the network and the transmission of raw biosignals or audio only for policy selection.

## IV. EXPERIMENTAL METHODOLOGY

*A. Data and Model Evaluation*

The physiological corpus consists of the 15-participant WESAD benchmark [10] which are still commonly used in recent multimodal affect research [8, 9] and naturalistic recordings of 33 additional participants. The resulting corpus consists of 48 adults in home, workplace, commute and public space environments. The speech corpus includes 1,440 RAVDESS utterances, which are a benchmark that continues to be used in recent real-time SER studies [11] as well as 2,880 naturalistic VA commands from the same participants, resulting in 4,320 utterances in five target affect classes. The Android app and the experience-sampling prompts were employed for collecting context and self-reported valence/arousal.

To avoid identity leakage, data was partitioned 70% training, 15% validation and 15% held-out testing. The accuracy, precision, recall, macro F1 and AUC-ROC were used to evaluate the classification. Following recent explainability-guided SER analysis [12], we followed feature attribution with SHAP. The RL environment was calibrated based on a diary study of 1 week prior to deployment and 100,000 simulated episodes were produced of common and rare affect-context combinations. To test convergence stability, five runs were made with seed numbers set as random. The convergence criterion was a 2% difference in the rolling average reward over 1000 episodes of training. The field deployment, which used the simulation for design, gathered feedback to be analyzed after a specific policy was implemented, but not in the sense of allowing unsafe online exploration.

Self-reports of valence/arousal, experimentally labelled benchmark data and temporally close physiological or behavioural events were used to achieve ground truth alignment. Only short gaps were filled in with interpolation of missing sensor samples, and low quality sensor segments due to a signal to noise ratio threshold were discarded. To minimize majority-class and identity effects, class balancing and stratification of participants was employed. The naturalistic extension adds to the ecological validity but doesn't remove the label noise that is found in self-reported emotion.

*B. Field Study*

A within-subject crossover study was conducted over 6 weeks that included three conditions: baseline assistant, context-only assistant, and EmpathicVA. The 48 participants were well-balanced in terms of gender (24 women, 24 men), and were an average of 32.4 years (SD=8.7) old, and reported at least three VA interactions per week. Order counterbalanced, baseline measurement and sensor calibration in week 1, followed by two week-long experimental periods, separated by one week baseline washout.

Outcomes measured were five-point satisfaction ratings each day, timing appropriateness, perceived emotional understanding, trust, perceived intrusiveness, weekly System Usability Scale (SUS) scores, interaction logs and physiological changes associated with interruptions. Exit interviews were independently coded by two researchers. Response latency, task completion, dismissal, command complexity and abandonment were recorded in interaction logs. Physiological responses were measured for the 30 s immediately following each interruption compared to the 30 s preceding the interruption. Raw audio and biosignals were processed locally, with only derived feature vectors sent. Participants gave their individual permissions for the physiological, audio, location and calendar streams and could opt out of having the data stored.

A crossover design minimizes between-person sources of variability, but may present carryover effects. These were restricted by using counterbalancing and the washout week. Within the participant, comparisons were made using statistical analyses and effect sizes were reported along with the significance tests to differentiate practically meaningful changes and small changes that may be significant with repeated measurement.

*C. Outcome Definitions*

Unnecessary interruption was defined as an assistant intervention in which a child was interrupted before it was necessary, the child ignored or dismissed the intervention, it was quickly silenced, or there was a noticeable physiological stress reaction that did not result in task engagement. Five-point scales were used for satisfaction, appropriateness of timing, emotional understanding, trust, and intrusiveness. The standard 0–100 scale of the SUS was used for analysis.

The primary comparison was EmpathicVA versus baseline; context-only served as an ablation of affective sensing. This design separates gains attributable to ordinary situational awareness from gains requiring physiological and vocal affect. Qualitative themes were retained only when supported by repeated coded evidence. Inter-rater reliability was measured using Cohen's $\kappa$, and disagreements were resolved through discussion. The combination of self-report, behavior, and physiology was used to reduce dependence on any single evaluation channel.

TABLE II
PERFORMANCE BY EMOTION CATEGORY

| Emotion | Precision | Recall | F1 | Support |
|---|---|---|---|---|
| Neutral | .945 | .952 | .948 | 2,847 |
| Positive | .912 | .898 | .905 | 1,923 |
| Negative | .889 | .907 | .898 | 1,654 |
| Stressed | .931 | .945 | .938 | 2,156 |
| Excited | .876 | .863 | .869 | 1,205 |
| Macro average | .911 | .913 | .912 | 9,785 |

TABLE III HELD-OUT AFFECT RECOGNITION PERFORMANCE

| Modality | Acc. | Prec. | Recall | F1 / AUC |
|---|---|---|---|---|
| Physiological | .863 | .858 | .863 | .860 / .941 |
| Vocal | .847 | .852 | .847 | .849 / .928 |
| Context | .612 | .598 | .612 | .605 / .789 |
| Multimodal fusion | .923 | .921 | .923 | .922 / .967 |

## V. RESULTS

### *A. Affect Recognition and Policy Learning*

Table III shows that multimodal fusion achieved 92.3% accuracy, 0.922 macro F1, and 0.967 AUC-ROC. It improved accuracy by 6.0 points over the physiological-only model, 7.6 over vocal-only, and 31.1 over context-only. Neutral and stressed states were the most reliably recognized (F1=0.948 and 0.938), while excited was the most difficult (F1=0.869), primarily because high arousal overlaps with positive and stressed states. Prediction consistency across consecutive windows was 87.3%, transition smoothness was 92.1%, and mean stable-detection latency after a genuine state change was 12.4 s. Errors were concentrated between affectively adjacent classes rather than across opposite valence-arousal regions. In particular, negative was sometimes confused with stressed, and excited with positive or stressed, which is consistent with their shared arousal characteristics.

SHAP analysis ranked HRV RMSSD, mean fundamental frequency, phasic EDA, and the first MFCC as the strongest features. Contextual variables collectively contributed 21.2% of total importance, confirming that context provides nonredundant disambiguation rather than merely duplicating physiological or vocal evidence.

The Double DQN converged after $6{,}847 \pm 432$ episodes across five runs, with final average reward $0.847 \pm 0.023$ and 94.7% action consistency in the final 1,000 episodes. The learned policy was interpretable: immediate response dominated positive (62.3%) and excited (58.9%) states, whereas stressed states shifted toward extended delay (24.7%), empathetic response (29.4%), and silent mode (18.0%). Immediate responses fell to 12.3% during stress. Negative affect produced the highest empathetic-response rate (31.2%). Table IV reports the complete action distribution. The policy did not map every negative or stressed state to silence; it balanced message urgency, recent user feedback, and uncertainty. This is important because a rigid "never interrupt when stressed" rule would suppress urgent or explicitly requested assistance.

TABLE IV
LEARNED ACTION DISTRIBUTION BY AFFECT (%)

| State | Imm. | Brief | Ext. | Emp. | Silent |
|---|---|---|---|---|---|
| Neutral | 45.2 | 28.4 | 12.1 | 8.7 | 5.6 |
| Positive | 62.3 | 19.8 | 7.2 | 6.9 | 3.8 |
| Negative | 18.7 | 23.4 | 19.8 | 31.2 | 6.9 |
| Stressed | 12.3 | 15.6 | 24.7 | 29.4 | 18.0 |
| Excited | 58.9 | 22.1 | 8.4 | 7.3 | 3.3 |

TABLE V
FIELD-STUDY OUTCOMES BY CONDITION

| Outcome | Baseline | Context-only | EmpathicVA |
|---|---|---|---|
| Satisfaction (1–5) | 3.24 | 3.78 | 4.51 |
| Timing appropriateness | 2.89 | 3.45 | 4.29 |
| Emotional understanding | 2.12 | 2.34 | 4.18 |
| Trust | 3.45 | 3.67 | 4.32 |
| Intrusiveness (lower better) | 3.78 | 3.22 | 2.14 |
| SUS (0–100) | 68.4 | 74.2 | 82.7 |
| Stress episodes | 89 | 42 | 12 |
| ΔHR during interruption (bpm) | +12.8 | +6.2 | +1.4 |
| ΔEDA during interruption ($\mu$S) | +2.4 | +1.1 | +0.3 |

### *B. User Experience and Physiological Outcomes*

EmpathicVA outperformed both comparison conditions across all primary subjective measures (Table V). Overall satisfaction increased from 3.24 for baseline to 4.51, timing appropriateness from 2.89 to 4.29, emotional understanding from 2.12 to 4.18, and trust from 3.45 to 4.32. Perceived intrusiveness decreased from 3.78 to 2.14. All reported comparisons were significant at $p < .001$, with absolute Cohen's $d$ values from 0.95 to 2.31. SUS improved from 68.4 (baseline) to 82.7.

Objective physiology showed the same pattern. Baseline interruptions increased heart rate by $12.8 \pm 4.3$ bpm, EDA by $2.4 \pm 0.8\mu$S, and reduced HRV RMSSD by $8.7 \pm 3.2$ ms. Corresponding changes under EmpathicVA were $+1.4 \pm 2.1$ bpm, $+0.3 \pm 0.4\mu$S, and $-0.8 \pm 1.9$ ms and were not statistically significant. Interruption-related stress episodes decreased from 89 to 12, an 86.5% reduction; unnecessary interruptions decreased by 73%.

Longitudinal logs further showed 47% more interactions, 23% longer sessions, 31% more multi-step commands, and 58% fewer abandoned interactions under EmpathicVA. Interview coding achieved $\kappa = 0.84$. The most common themes were personalization (92%), emotional validation (89%), increased trust (83%), and reduced cognitive load (75%). Privacy concern was reported by 31%, showing that improved experience does not remove concerns about continuous affect sensing. Behavioral logs were consistent with the interviews: 78% of participants shifted from terse commands toward more conversational phrasing, 64% used the system for at least one emotionally supportive interaction, and 89% showed greater tolerance of functional errors after the assistant had appropriately acknowledged their affective state. These observations should not be interpreted as clinical efficacy; they instead indicate a change in the perceived relationship between user and assistant.

### *C. Cross-Condition Interpretation*

The context-only assistant consistently occupied the middle position between baseline and EmpathicVA. It reduced stress episodes from 89 to 42 and improved SUS from 68.4 to 74.2, confirming that calendar, activity, location, and timing cues are useful even without emotion sensing. Nevertheless, emotional understanding changed only from 2.12 to 2.34, whereas EmpathicVA reached 4.18. The gap indicates that users distinguished ordinary context adaptation from behavior that appeared responsive to affect.

The largest subjective effect was emotional understanding ($d = 2.31$), followed by reduced intrusiveness ($d = -1.52$) and timing appropriateness ($d = 1.41$). Trust improved more moderately ($d = 0.95$), which is plausible because trust depends on reliability and privacy in addition to affective behavior. The physiological results strengthen the subjective findings: EmpathicVA's post-interruption changes were small and nonsignificant, whereas baseline produced clear sympathetic activation. Agreement across these measures reduces the likelihood that results are solely demand characteristics or novelty-driven positive ratings.

The policy's action distribution also provides a useful sanity check. It did not simply maximize silence. Positive and excited states received immediate responses in most cases, preserving responsiveness when interruption cost was low. During stress, the combined use of extended delay, empathetic response, and silent mode rose sharply. This state-dependent shift is the mechanism that connects affect recognition to observed reductions in unnecessary interruption.

## VI. Discussion

The results back up three messages. There is a timing dimension to empathetic interaction, first of all. Incorporating emotionally appropriate language is not enough in case of interruption under acute stress or high cognition load. This was partly because the learned policy incorporated more delay and silence in the stressful period, which is why the user ratings and physiological changes went up together.

Secondly, multimodal fusion is not just a statistic convenience, but an operational convenience. Physiological signals offer fairly direct arousal evidence, whereas vocal features offer a more transient expression of change, and context aids in the differentiation of otherwise ambiguous patterns. The advantage of the fusion model over unimodal alternatives, along with the coherent errors at the class level, suggest that the modalities provide complementary information.

Third, affective adaptation seems to be important for sustaining engagement and trust beyond the immediate satisfaction. The users' mental model of the assistant grew more competent and socially responsive, as the number of interactions and the complexity of the commands increased. That advantage, however, comes with a great deal of governance responsibility. The major risks that have been highlighted in recent reviews for conversational and emotion-recognition systems are privacy, trust, bias, autonomy, and inappropriate inference [14], [15]. Local processing, granular consent, visible explanations, revocation and data minimization should then be design features and not optional features.

There are some significant limitations to the study. The field deployment lasted for six weeks, included Android devices, and included mainly an educated sample of English speakers. The naturalistic extension is larger than original benchmark but still small for training high capacity multimodal models. While RL training was done in simulation, not with users, there is a possibility of simulation to field mismatch. The claimed benefits need to be replicated in different cultures, languages, devices, deployments and independently.

Another constraint is the need for wearable sensors. Continuous HRV and EDA enhance stress discrimination but at the cost of charging, comfort and adherence, which could hamper consumer uptake. When wearables are not available, voice and context need to be handled in a practical system through graceful degradation. On the other hand, voice-only may fail with no warning, in noisy environments, in shared spaces, or when the user intends to hide emotion. Thus, the architecture should rather consider modality availability and confidence as first-class inputs instead of the assumption of a fixed sensor set.

There is also a difference between detecting correlates of affect and knowing what a person is feeling. The five output categories make what is a continuous process, culturally mediated, easy to understand. EmpathicVA is not intended to make predictions that can be considered facts, or draw inferences about clinical conditions. It should be used to assess the risk of interactions and select a conservative action given uncertainty. The most convincing evidence in this study relates to interruption management, rather than emotional intelligence in general.

The results inspire the design principles for affect assessment: Use of complementary signals for assessing affect; Learned, statedependent policies for decisions; Acknowledging emotion when appropriate for engagement; Personalize timing over repeated interactions; and Transparency about inference and use of data.

*A. Deployment and ethical implications*

The physiological and behavioral signals add new risks with affect-aware systems, as the routines, vulnerability and wellbeing changes can be detected by these signals [14, 15]. Architectural level privacy protection should therefore be provided. This on-device feature extraction method minimizes exposure to the raw data, which can be sensitive, but may also contain information derived from it that might also be sensitive. Deployed systems should: a) minimize retention, b) break consent by modality, c) make clear when affect inference is active, d) make understandable reasons why inferences are delayed or not made, and e) allow users to disable or correct inference. A policy such as the EU Artificial Intelligence Act [16] has a transparency, human oversight and risk management focus, which aligns with these controls. Training goals should not maximize engagement with minimal autonomy, otherwise, emotionally manipulative behavior may be learned.

There's a question of fairness, too. There are differences in physiological baselines, prosody of voice, norms for the display of emotions, and attitudes toward interruption across individuals and cultures. While some physiological variation has been minimized, subject normalization does not ensure equal performance. Prior to deployment, it is important to audit the performance of each demographic and linguistic group, to surface any uncertainty into the policy and to use conservative fallback behavior if the model is poorly calibrated.

*B. Future Work*

The effects of policy need to be measured over longer periods to assess whether post novelty effects exist and whether personalization fades as routines evolve. The affect classifier as well as social appropriateness of learned actions should be validated across cultures and languages. Federated training and differential privacy are examples of privacy-preserving learning techniques that may minimize the impact of central exposure, but their impact on model calibration needs to be quantified. Finally, independent replication is required since the reported naturalistic corpus and deployment were generated in a single study pipeline.

*C. Practical Design Guidance*

The findings suggest some implementation criteria for future AWE based assistants. First, the quality of sensing needs to be brought into the decision policy. When the data is not available (such as a calendar), or the speech is disorganized or noisy, the confidence should decrease and so should the emotional labeling, not the timing. Second, urgency should not be correlated with feeling. Safety alerts, and user-chosen requests, should be delivered even if stress is detected; affect may change the wording and timing as long as it does not exceed the limits set by the message priority.

Third, personalisation needs to be reversible. A policy that adapts to a user's habits may become inappropriate after job changes, traveling, illness, or other changes in the household environment. Systems must thus be made to have explicit reset controls, easily visible preference histories, and regular confirmation rather than the old behavior being assumed to be consent. Fourth, emotional acknowledgment should be short and commensurate. Over-stating and repeating comments about how a user feels can be patronizing and can even lead to the showing of inferences in shared areas. Many times, it is more considerate to hold back from reacting emotionally rather than reacting at all.

Lastly, a measure of evaluation should not be limited to classification accuracy. Even with a good affect classifier, the output of the assistant may be sub-optimal, if the errors in the affect classifier are in high-cost situations, or if the policy reacts too intensely to uncertain situations. Calibration, false stress alarms, missed periods of stress, acceptance of interruptions, stress costs, task completion, and user control are all key measures. The evaluation here is therefore more robust than a benchmark only comparison but further replication and sampling are required.

## VII. Conclusion

EmpathicVA combines multiple modalities of affect recognition and an interruption policy for voice assistants based on RL. The system was able to reach an affect-recognition rate of 92.3% on held-out data and to train a meaningful timing strategy that encourages engagement in the receptive state and inhibits it in the stress state. The framework was found to increase satisfaction, usability, trust and timing appropriateness, and to decrease intrusiveness and physiological stress in a six-week study. The results indicate that the EEAs should have both the capacity to perceive accurately and to not interrupt in a disciplined way. Future research should focus on longterm adaptation, cross-cultural validation, privacy preserving learning and independent replication.